\documentclass[printer]{aamod}
\usepackage{graphicx}
\usepackage[varg]{txfonts}
\usepackage[colorlinks]{hyperref}
\usepackage{natbib}
\bibpunct{(}{)}{;}{a}{}{,} 
\usepackage{revcomments}
\usepackage{lscape}

\begin{document}

\title{Modelling solar low-lying cool loops with optically thick
radiative losses.}
\titlerunning{Modelling solar low-lying cool loops with optically thick radiative losses.}
\author{C. Sasso\inst{1} \and V. Andretta\inst{1} \and D. Spadaro\inst{2}}

\offprints{C. Sasso, \email{csasso@oacn.inaf.it}}

\institute{INAF-Osservatorio Astronomico di Capodimonte, Salita Moiariello 16,
I-80131 Napoli, Italy \and INAF-Osservatorio Astrofisico di Catania, Via S. 
Sofia 78, I-95123 Catania, Italy}

\date{Received / Accepted}

\abstract{}{We investigate the increase of the DEM (differential emission
  measure) towards the chromosphere due to small and cool magnetic loops
  (height $\lesssim8$~Mm, $T\lesssim10^5$~K). In a previous paper we analysed
  the conditions of existence and stability of these loops through
  hydrodynamic simulations, focusing on their dependence on the details of the
  optically thin radiative loss function used.}{In this paper, we extend those
  hydrodynamic simulations to verify if this class of loops exists and it is
  stable when using an optically thick radiative loss function. We study 
  two cases: constant background heating and a heating depending on the
  density. The contribution to the transition region EUV output of these loops
  is also calculated and presented.}{We find that stable, quasi-static cool
  loops can be obtained by using an optically thick radiative loss function
  and a background heating depending on the density. The DEMs of these loops,
  however, fail to reproduce the observed DEM for temperatures between
  $4.6<\log T<4.8$. We also show the transient phase of a dynamic loop
  obtained by considering constant heating rate and find that its average DEM,
  interpreted as a set of evolving dynamic loops, reproduces quite well the
  observed DEM.}{} 


\maketitle

\section{Introduction}

The origin of the EUV output at temperatures below 1~MK is still widely
debated in Solar Physics. The classical picture that the transition region
(hereafter TR) emission originates from the base of the hot large-scale
coronal loops strongly underestimates the observed EUV emission below 0.1 MK,
but no alternative, quantitative view has gained consensus to-date. One of the
proposed explanations hypothesizes that much of the TR plasma is confined in
relatively small and cool magnetic loops (height $\lesssim8$~Mm,
$T\lesssim10^5$~K), which are directly connected to the chromosphere but
thermally insulated from the corona \citep{dowdy1,dowdy2,feldman1,feldman4}.
 
From an observational point of view, these loops are indeed very difficult to
observe. The first, presumed direct observations present in the literature 
have been obtained with the VAULT instrument \citep[Very High Angular
Ultraviolet Telescope,][]{vault} in the \ion{H}{i} Ly-$\alpha$ line. They show
loop-like structures with estimated temperatures and densities
($T=10^4-3\times10^4$~K, $P=0.1-0.3$~dyne~cm$^{-2}$) that could be appropriate
for the low-temperature end of cool loops \citep{patsourakos,vourlidas}. This
interpretation has been debated by \citet{judge}. More recently, the
launch of the IRIS spacecraft \citep{IRIS}, in June 2013, has given new
possibilities to observe these loops. The analysis of the data obtained in
spectral lines and continua covering a range of temperatures $\log T=3.7-7$~K
with a spatial resolution of $\sim 0.4$", represents a very good opportunity
to look for structures with the dimension and temperatures of the class of
loops described above. It is therefore not surprising that observations of
highly dynamical cool, low-lying loops, in many respects similar to those we
discuss in this paper, have recently been reported \citep{hansteen}.

In a previous paper \citep[][hereafter referred to as Paper I]{sasso}, we 
analyzed the general properties of quasi-static (velocity along the loop lower
than 1~km/s) cool loops with $T\lesssim0.1$~MK and their conditions of
stability and existence under different and more realistic assumptions on the
optically thin radiative loss function with respect to previous works
\citep[i.e.,][]{cally}. In particular, we obtained through hydrodynamic
simulations stable low-lying cool loops, even for a set of parameters that
would prevent the formation of rigorously static loops. The existence of the
loops we found is due indeed to small departures from static conditions,
i.e. to the presence of a small but non-zero conductive flux and velocities,
and to the requirement of nearly constant pressure (implying that our loops
are limited to low heights above the chromosphere). In our simulations, we
considered only the case of constant heating rate. We also showed that the
emission of these cool loops, plus the emission of intermediate temperature
loops ($0.1<T<1$~MK), can account for the observed radiative output below 1~MK. 

From a theoretical point of view, there are still several points that need to 
be explored in order to determine the conditions under which cool loops could
exist in the solar atmosphere. One important point is the shape of the
radiative loss function below 0.1~MK, due to the presence of the \ion{H}{i} 
Ly-$\alpha$ peak, which is very important for the existence of cool loops. 
  
Our work is based on 1-D hydrodynamic simulations and aims at studying the
conditions of existence of cool loops to understand, in particular, the
mechanisms of their heating and energy balance through comparison between
their simulated differential emission measure (hereafter, DEM) and the
observed one. \citet{peter,dem} made the first successful attempt to reproduce
the shape of the DEM curve quantitatively and qualitatively, even at
temperatures below $\log T=5.3$~K. They synthesized spectra from
three-dimensional MHD simulations of the whole Sun atmosphere, finding
structures that could be related to the kind of loops we are
studying. However, the cool loops we describe would be covered by only very
few resolution elements in their simulation, and in any case resolving the
gradients and the dynamics of the relevant quantities in our loop models would
require a much higher resolution. Therefore, we regard our study as
complementary to large-scale 3-D simulations.

As in Paper I, while looking for cool loops, we have also found low-lying
quasi-static loops with temperatures in the range $1-5\times
10^5$~K. Following one of the latest loop classifications \citep{realesolo1},
we should refer to these loops also as ``cool coronal loops''. In order to
avoid confusion, we will refer to them as ``intermediate-temperature loops''.

In this paper we want to make a further step in the direction of considering
more realistic assumptions for the simulations of cool loops with respect to
Paper I, by introducing an optically thick radiative loss function. In
Sec.~\ref{sec:model}, we describe the numerical model and introduce the
radiative loss function adopted. In Sec.~\ref{sec:results}, we present the
hydrodynamic simulations and the loops obtained (cool and
intermediate-temperature loops) with different assumption on the heating rate
and we discuss and analyze their properties. Section~\ref{sec:dem} is
dedicated to the calculated DEMs of the loops obtained and to the comparison
with the observed one. Finally, in the conclusions (Sec.~\ref{sec:concl}), the
role of the cool and intermediate-temperature loops in the solar atmosphere
and the comparison with the observations is treated. 

\section{Numerical calculations}\label{sec:model}

The set of hydrodynamic equations for mass, momentum, and plasma energy
conservation for a fully ionized hydrogen plasma have been solved in a
unidimensional, magnetically confined loop of constant cross-section with
ARGOS, a 1-D hydrodynamic code with the fully adaptive-grid package PARAMESH
\citep{ARGOS,PARAMESH}. A fully adaptive-grid is necessary to adequately
resolve one or more evolving regions of steep gradients. The hydrodynamic
equations for mass, momentum, and energy, respectively, solved by ARGOS are
\begin{eqnarray}
\frac{\partial}{\partial t}\rho+\frac{\partial}{\partial s}(\rho v)&=&0, \label{eq:1}\\
\frac{\partial}{\partial t}(\rho v)+\frac{\partial}{\partial s}(P+\rho v^2)&=&-\rho g_{\parallel}(s), \label{eq:2}\\
\frac{\partial U}{\partial t}+\frac{\partial}{\partial s}\left(Uv+F_{\textmd{\tiny{c}}}\right)&=&-P\frac{\partial}{\partial s}v+E(s,t)-n^2\Lambda(T,P), \label{eq:3}\\
F_{\textmd{\tiny{c}}}&=&-10^{-6}T^{5/2}\frac{\partial}{\partial s}T.\label{eq:condflux}
\end{eqnarray}
where $t$ is the time, $\rho$ the mass density, $v$ the velocity, $P$, $T$ and
$n$ are the gas pressure, temperature, and electron number density,
respectively. $U$ is the internal energy, $s$ the curvilinear coordinate
along the loop, $E(s,t)$ the assumed form for the input heating rate,
$n^2\Lambda(T,P)$ the plasma radiative losses specified by the radiative loss
function $\Lambda(T,P)$, $g_{\parallel}(s)$ the component of the solar gravity
along the loop axis, and $F_{\textmd{\tiny{c}}}$ the thermal conductive flux,
in CGS units.

The code is based on a loop geometry that assumes an arched loop of a given
length $L$ and apex height above the chromosphere $h$ as described in 
\citet{karpen,spadaro}. At each footpoint of the loop there is a thick
chromosphere (26.7~Mm deep) acting as a mass reservoir, with temperature set
to $T=9.5\times10^3$~K. Since we take, by definition, the top of the
chromosphere as the level at which the plasma drops below $9.5\times10^3$~K,
the exact position of the top of the chromosphere ($s=\pm L_i/2$ at the
beginning of the simulation, $s$ being the curvilinear coordinate along the
field lines) changes during the calculation with the plasma filling or
evacuating the loop. So, at end of the simulation, we will have a new position
for the top of the chromosphere $s=\pm L_f/2$ and, consequently, a new value
of $h=h_f$ , where $h_f$ is no longer the geometrical parameter defining the
shape of the loop, but the height of the loop apex above the $T=9.5\times10^3$~K level. 

The main input parameters for the calculations are the radiative loss
function, the heating rate, the pressure (or the density) at the chromospheric
reference temperature, and the loop geometry ($h$ and $L$). In Paper~I, we
used constant heating rates per unit volume throughout the loop. Following the
more general approach of \cite{an86}, we also consider the case of a constant
heating rate per particle. The two cases are parametrized as follows:
\begin{equation}
E(s,t) = E_\mathrm{h}\: f(s)\: [ n(s,t)/n_\circ ]^\gamma \, , \label{eq:heating}
\end{equation}
where $\gamma=0$ is the case of constant heating per unit volume, and $\gamma
= 1$ corresponds to the case of constant heating per particle and $n_\circ =
3.9882 \times 10^9$~cm$^{-3}$ is the value of the density at the base of the
loop, taken from the work of \citet{kuin}. The function $f(s)$ specifies the
variability of the heating rate (per particle or per volume) along the loop.
With the exception of the discussion of Sec.~\ref{sec:notes}, we will assume
$f(s)=1$ throughout this paper. The radiative loss function adopted in this
paper is described more in detail in the following section.

\subsection{Radiative loss function}\label{sec:radlosses}

In order to estimate the radiative losses in the optically thick H Ly-$\alpha$
line, we used the calculations by \citet{kuin}. Those authors computed the
contribution to radiative losses of hydrogen and helium taking into account
the effects of geometry and optical depths and the non-LTE (non-local
thermodynamic equilibrium) ionization state of hydrogen and helium. They
generated 3-D tables of radiative losses as a function of $T$, $P$ and slab
thickness, for H and He. We combine those tables with the radiative losses of
the other elements from the CHIANTI database \citep[version
7.1,][]{landi} and the code interpolates these tables depending on the
temperature and the pressure of the loop. We will refer to the resulting
radiative loss function as $\Lambda_\mathrm{kp}$. We consider in the following
calculations only the case of slab thickness equal to 200~km. This value is
consistent with the dimensions of these loops as inferred from observations
\citep{vourlidas,hansteen}.

There are other calculations of optically thick radiative losses in the
literature, like the work of \citet{carlsson} that is indeed more recent. This 
work is dedicated to the radiative cooling and heating only in the
chromosphere, by combining detailed non-LTE radiative transfer calculations
and time-dependent 2D MHD simulations. We decided to use the radiative losses
calculated by \citet{kuin}, even if older, because their results are presented
in a form that can be easily incorporated in hydrodynamic flux-tube
calculations and are expressly aimed at flux-tube modelling. In addition,
their calculations are relevant to a broader temperature range more suitable
for our calculations.   

\begin{figure}
\centering
\includegraphics[clip=true,angle=90,width=9cm]{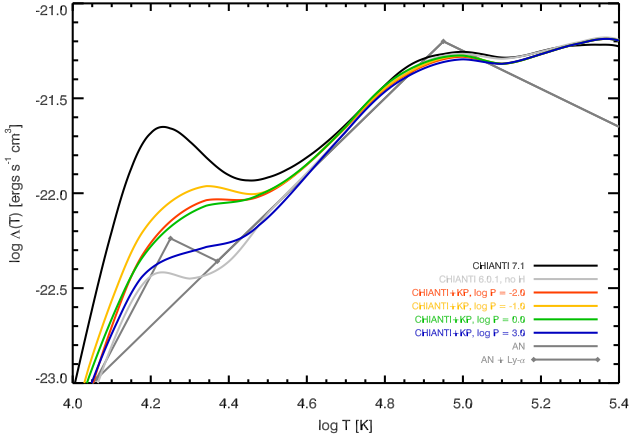}
\caption{Radiative loss function $\Lambda_\mathrm{kp}$ plotted for different
  pressure values ($\log P$= -2, -1, 0, 3, red, yellow, green and blue line,
  respectively) and the one from the CHIANTI database, version 7.1
  \citep{landi} (black line). Moreover, we plot some of the radiative loss
  functions used in Paper I: \citet[AN, dark grey line,][]{an86}; AN function
  plus a peak mimicking the H Ly-$\alpha$ losses (dark grey line plus diamond
  symbols); from the CHIANTI database, version 6 \citep{chianti}, without the
  H contribution (light grey line).}
\label{fig:1}
\end{figure}

In Fig.~\ref{fig:1} we show the radiative loss function $\Lambda_\mathrm{kp}$
plotted for different pressure values ($\log P$= -2, -1, 0, 3, red, yellow,
green and blue line, respectively) and the one from the CHIANTI database,
version 7.1 \citep{landi} (black line) from which we start to compute them. They
are compared to some of the radiative loss functions used in Paper I, for
which we obtained stable cool loops, that are: power-law segments function
equal to $T^2$ for $\log T<4.95$~K and $T^{-1}$ for $\log T>4.95$~K \citep[AN,
dark grey line,][]{an86}; AN function plus a peak mimicking the H Ly-$\alpha$
losses (dark grey line plus diamond symbols); from the work of \citet{chianti}
without the H contribution (light grey line). The black and the blue line
represent the upper and the lower limit for the radiative loss functions:
optically thin and optically thick case, respectively.  

\subsection{Preliminary considerations on cool loop solutions}\label{sec:notes}

In their work, \citet{an86} turned their attention to the solutions of the
hydrodynamic equations of loops with negligible conductive flux, and studied
the properties and conditions of existence of their solutions under specific
hypotheses about the radiative loss functions. They in particular approximated
the optically thin function $\Lambda(T)$ with power-law segments: $\Lambda(T)
\sim T^a$ for $T< 0.1$~MK, and $\Lambda(T) \sim T^{-b}$ for $T > 0.1$~MK, with
$a$ and $b$ positive values. The conditions of existence and stability of such
solutions were further and extensively studied by, e.g.,
\citet{klimchuk,cally}. Here we revisit some of those previous analyses,
extending the results to consider the specific radiation losses functions we
used in our simulations.

\citet{an86} solved the hydrodynamic equations Eq.\ref{eq:1}--\ref{eq:3}
assuming negligible conductive flux. The energy equation, Eq.~\ref{eq:3}, can
then be rewritten as:
\begin{equation}
\left(\frac{P}{2kT}\right)^2\Lambda(T,P)=E(s,P) \; , \label{eq:fczero}
\end{equation}
where now we considered the more general case of optically thick radiative
loss function, $\Lambda(T,P)$, while $E(s,P)$ is given by
Eq.~\ref{eq:heating}.

It is convenient to define the following quantities:
\begin{eqnarray}
H(T) & \equiv & 2kT/(m_\mathrm{H}\,g) \; , \label{eq:H_scale} \\
\eta(s) & \equiv & g_\parallel(s)/g \; , \label{eq:g_parallel} \\
a(T,P) & \equiv & \partial\log\Lambda(T,P)/\partial\log T  \; , \label{eq:aT} \\ 
b(T,P) & \equiv & \partial\log\Lambda(T,P)/\partial\log P  \; . \label{eq:BT}
\end{eqnarray}
The quantity $a(T,P)$ can be interpreted as the local power-law index of the
radiative loss function at a given temperature and pressure. The values of
$a(T,P)$ for $\Lambda_\mathrm{kp}$ in the interval $\log T=$4.3--5 range from
$\approx 0.5$ to $\approx 2$. The values for $b(T,P)$ are substantially
smaller, ranging from $\approx-0.15$ to $\approx 0.2$ around the peak
temperature of the Ly-$\alpha$ line.

Substituting Eq.~\ref{eq:fczero} and Eq.~\ref{eq:heating} into the momentum
equation, Eq.~\ref{eq:2}, with the above definitions the the equation for $T$
becomes:
\begin{equation}
\left[a(T,P)+\gamma-2\right] \frac{\mathrm{d}T}{\mathrm{d}s} = 
T \frac{1}{f(s)}\frac{\mathrm{d}f(s)}{\mathrm{d}s} +
\left[b(T,P)+2-\gamma\right] \frac{T_\circ}{H(T_\circ)} \eta(s) \; , \label{eq:gradT}
\end{equation}
where $T_\circ$ is the temperature at the lower boundary of the loop.

The special case $a(T,P)=$ constant $=2-\gamma$ (power-law dependence of
$\Lambda(T,P)$ with temperature, with exponent either 2 or 1, depending on the
value of $\gamma$), reduces the above differential equation to the simpler 
expression:
\begin{displaymath}
H(T) = \left[-\frac{1}{f(s)}\frac{\mathrm{d}f(s)}{\mathrm{d}s}\right]^{-1} 
\left[b(T,P)+2-\gamma\right]\:\eta(s) \; .
\end{displaymath}
In this case, any function $f(s)$ that is monotonically decreasing with height
produces a loop solution, provided that $b(T,P)>-2+\gamma$ (true for
$\Lambda_\mathrm{kp}$).  The case $b=0$ and $a=2$ is the case we labelled
``AN'', and is shown in Fig.~\ref{fig:1} with a grey line.

In the remainder, we consider only the case $f(s)=1$; for simplicity, we
further neglect the dependence of $\Lambda$ on pressure, i.e.: $b(T,P) =
0$. In this case, Eq~\ref{eq:gradT} can be integrated to obtain an implicit
dependence of $T$ on $s$:
\begin{equation}
\theta(T,T_\circ) = 
\frac{1}{H_\circ} \int_{s_\circ}^s \eta(\sigma)\:\mathrm{d}\sigma \; , 
\label{eq:gradT_abar}
\end{equation}
where the function $\theta(T,T_\circ)$ is defined as:
\begin{equation}
\theta(T,T_\circ) \equiv 
\left[\frac{\bar{a}(T,T_\circ)+\gamma-2}{2-\gamma}\right]
\left(\frac{T}{T_\circ}-1\right) \; , \label{eq:theta_T}
\end{equation}
while the quantity $\bar{a}(T,T_\circ)$ is the ``mean'' power-law index:
\begin{equation}
\bar{a}(T,T_\circ) \equiv 
\frac{1}{T-T_\circ} \int_{T_\circ}^T a(\tau)\:\mathrm{d}\tau
\label{eq:abar_T} \; .
\end{equation}
The above equations highlights a first constraint for the existence of this
kind of solutions: $\bar{a}(T,T_\circ) > 2-\gamma$.  We have mentioned before
that for the radiative loss function we are using, we have $a<2$ for $\log
T>4.3$, it is clear that it would be very difficult to obtain cool solutions
for the case of uniform heating per unit volume, $\gamma=0$.  

\begin{figure*}
\centering
\sidecaption
\includegraphics[clip=true,angle=90,width=12cm]{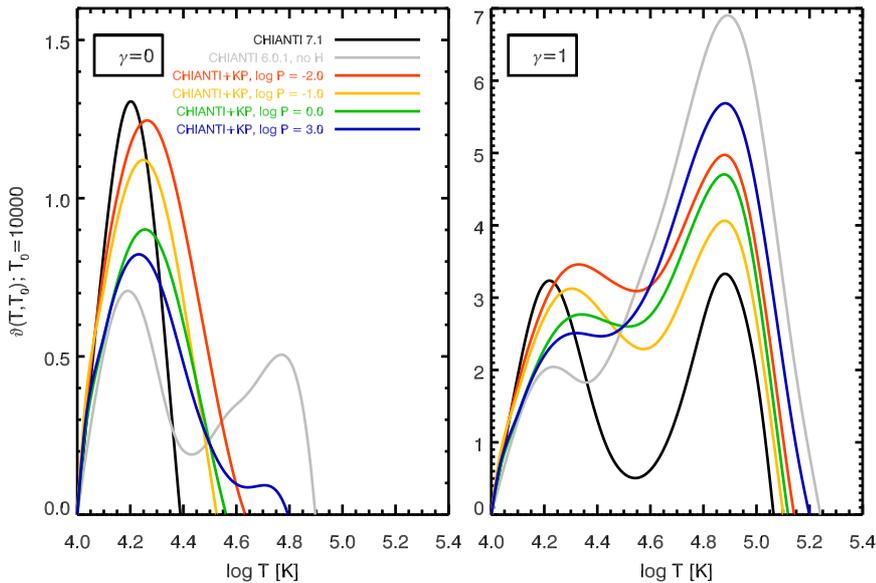}
\caption{Function $\theta(T,T_\circ)$ for the radiative loss functions shown
  in Fig.~\ref{fig:1} in the case of $\gamma=0$ (left-hand panel) and
  $\gamma=1$ (right-hand panel).}
\label{fig:2}
\end{figure*}

The upper limit to the loop temperature as mapped by function
$\theta(T,T_\circ)$ is given by the maximum of $\int_{s_\circ}^s
\eta(\sigma)\:\mathrm{d}\sigma/H_\circ$; in the case of a semicircular loop of
radius $h$, this is $h/H_\circ$.  However, a stronger constraint obviously
follows from the consideration that $\int_{s_\circ}^s
\eta(\sigma)\:\mathrm{d}\sigma$ is a monotonically increasing function,
whereas $\theta(T,T_\circ)$ is not, in general.  Single-value solutions are
therefore limited to the first local maximum of $\theta(T,T_\circ)$, shown in
Fig.~\ref{fig:2} for the radiative loss function of Fig.~\ref{fig:1} in the
case $\gamma=0$ and $\gamma=1$. 
  
The above simple considerations highlight one of the basic characteristics of
cool loop solutions: their strong sensitivity on the details of the heating
and of the radiation loss function.

\section{Results and discussion}\label{sec:results}

We ran numerous simulations, extensively exploring the parameter space, under
different initial conditions. We consider a loop in a
quasi-static equilibrium state when the plasma velocities are lower than
1–2~km/s. We took as an initial equilibrium state ($t=0$~s) for new
simulations some of the cool loops obtained in Paper~I, only changing the
heating rate and the radiative loss function. The list of simulated loops,
together with the relevant parameters, is given in Table~\ref{tab:1} (cool
loops) and Table~\ref{tab:2} (intermediate temperature loops).
 
\subsection{Loops from spatially uniform and temporally constant 
heating rate per unit volume}\label{sec:heatingvolume}

We start by making simulations with constant heating rate ($\gamma=0$) and
using the radiative loss function $\Lambda_\mathrm{kp}$. As expected from the
discussion in Sec.~\ref{sec:notes}, we are not able to obtain stable cool
loops since during the simulations, they become all hot ($T\sim
8\times10^5$~K). 

\begin{figure*}
\centering
\sidecaption
\includegraphics[clip=true,angle=90,width=12cm]{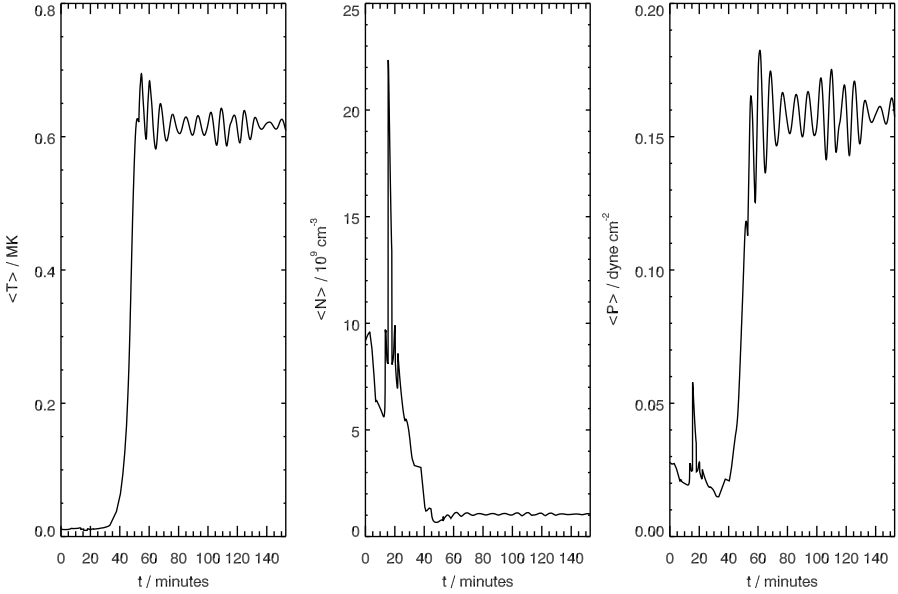}
\caption{Evolution of the mean temperature, mean density, and mean pressure of 
Loop0, obtained from a simulation starting from a stable cool loop of top
temperature $\sim1.2\times10^4$~K and constant heating rate per unit volume.}
\label{fig:3}
\end{figure*}

As an example, Fig.~\ref{fig:3} shows the evolution of the
mean temperature, density, and pressure of a loop (hereafter, Loop0) during a
simulation started from a stable cool loop of top temperature
$\sim1.2\times10^4$~K (Loop 24 of Table~1 in Paper I), assuming constant
heating rate.  

Loop0 stays for $\sim 42$~min in a cool state ($T< 10^5$~K) even if not a
stable one. The mean temperature of the loop oscillates between $1-2\times
10^4$~K (see left panel of Fig.~\ref{fig:3}) for $\sim 38$~min and then in
$\sim 4$~min reaches much higher values. It becomes a quasi-static
“coronal” loop, after $\sim 2.5$~h from the beginning of the simulation, 
reaching a top temperature of $\sim 8.5 \times 10^5$~K. ARGOS gives the
possibility to follow the evolution of the loop by storing the loop's
parameters at previously defined time steps. During the 4 minutes mentioned
before, the simulation records three states characterized by maximum
temperatures of $\sim 5, 7$ and $9\times 10^4$~K, progressively. During the
evolution of the loop, the maximum temperature is not always localized at 
$s=0$ (loop center) but also along the loop, i.e. at different values of $s$.   

In Paper I, we obtained indeed quasi-static cool loops by using constant
heating rate but the radiative loss functions used were different. From
Fig.~\ref{fig:1}, it is clear that $\Lambda_\mathrm{kp}$ for pressure values
characteristic of cool loops ($\log P \sim -2$) is higher than the losses used
in Paper I and requires a higher thermal conductive flux from warmer regions
to be balanced. 

\subsection{Loops from constant heating rate per particle}

\begin{figure*}
\centering
\includegraphics[clip=true,trim=0  0 30 0,width=6.5cm]{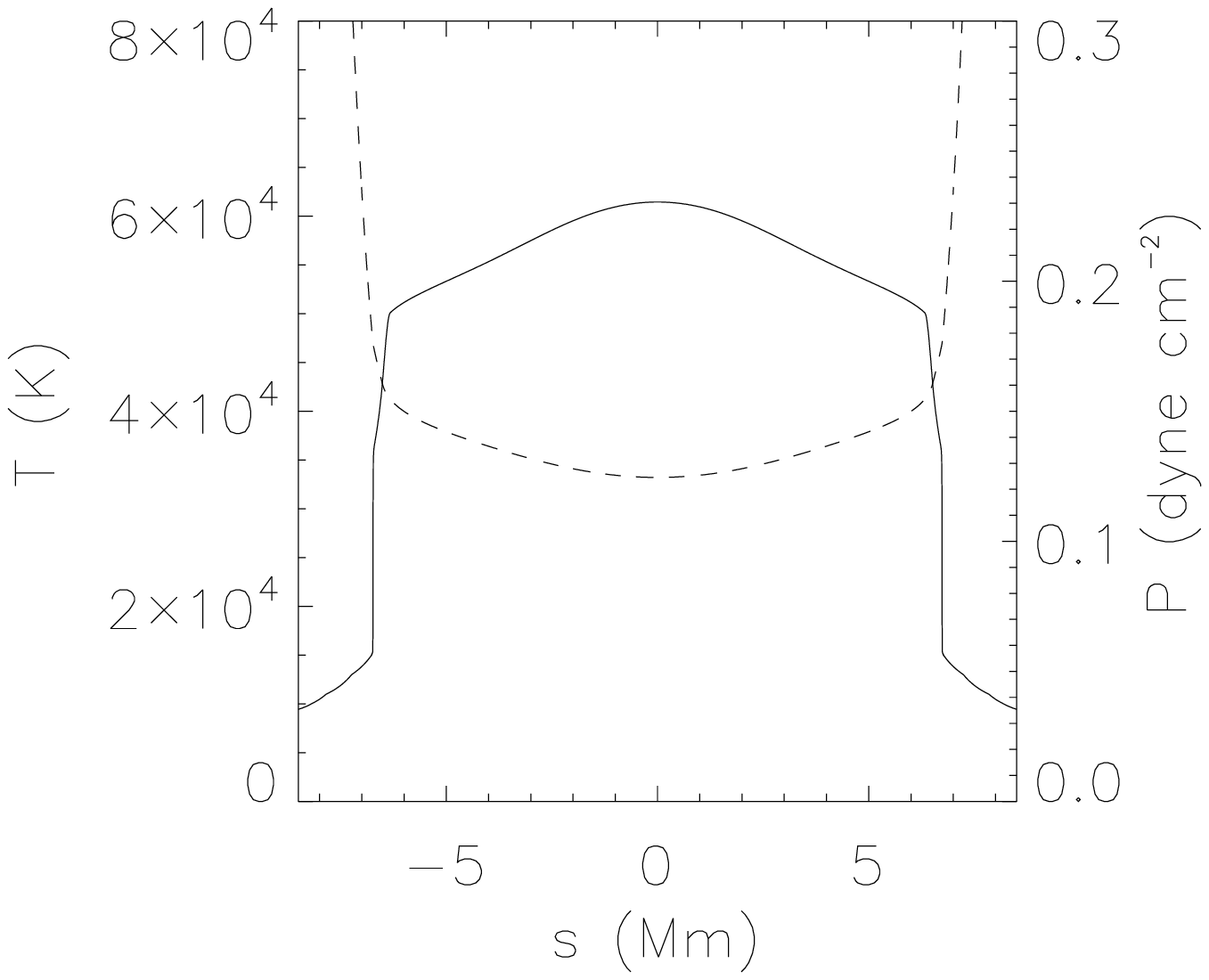}
\includegraphics[clip=true,trim=0  0 30 0,width=6.5cm]{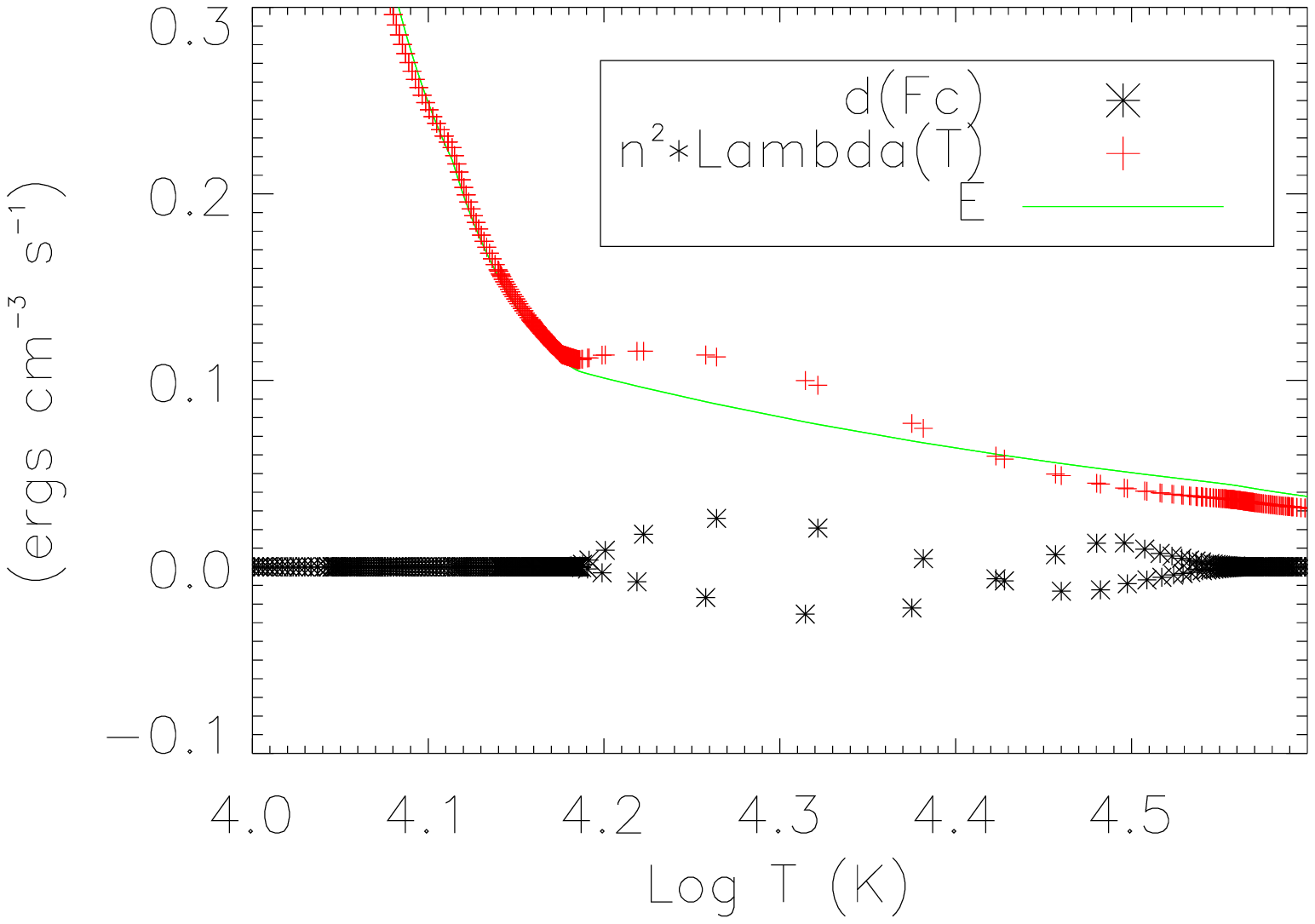}
\includegraphics[clip=true,trim=0  0 50 0,width=6.5cm]{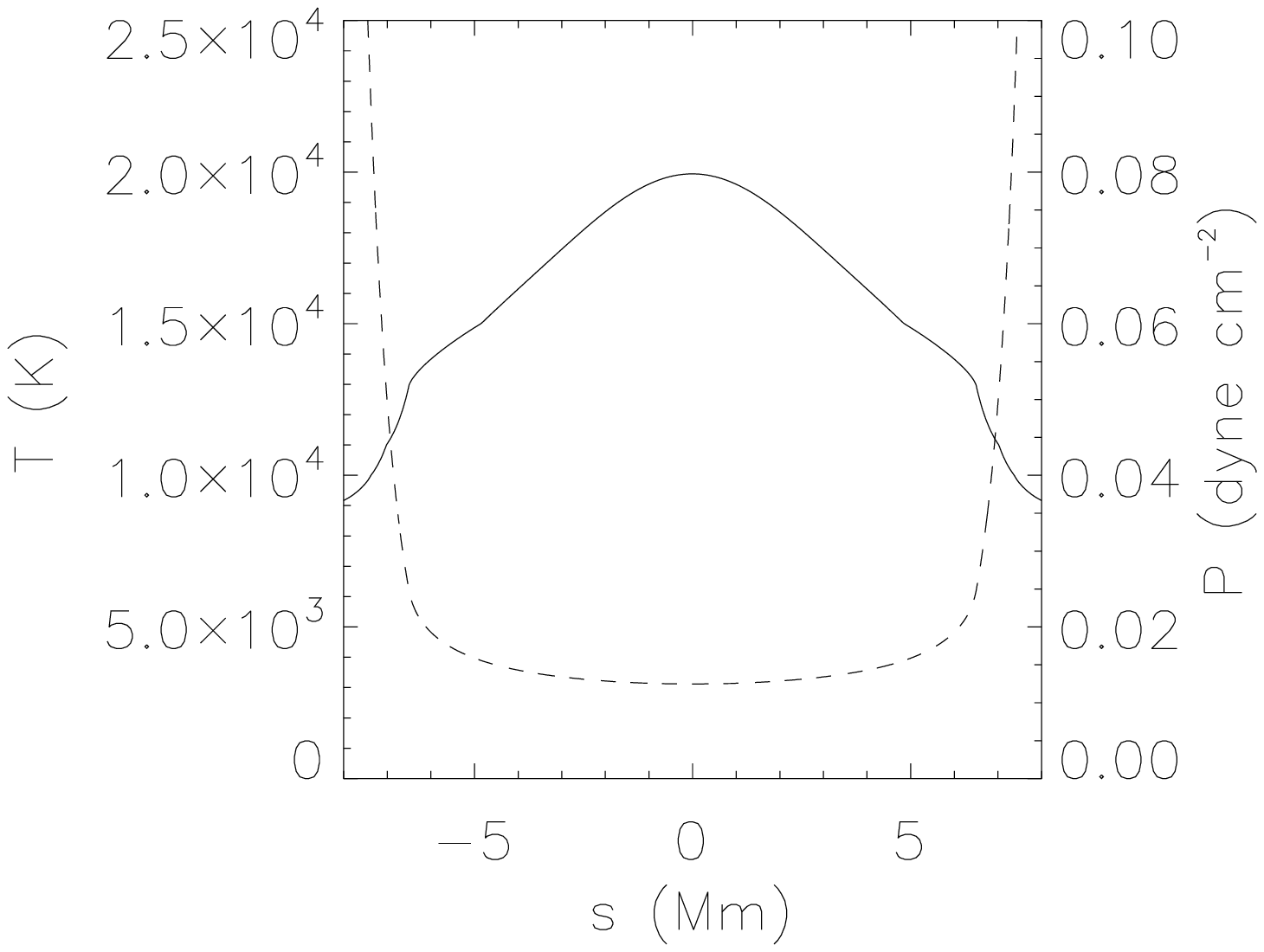}
\includegraphics[clip=true,trim=-5  0 30 0,width=6.5cm]{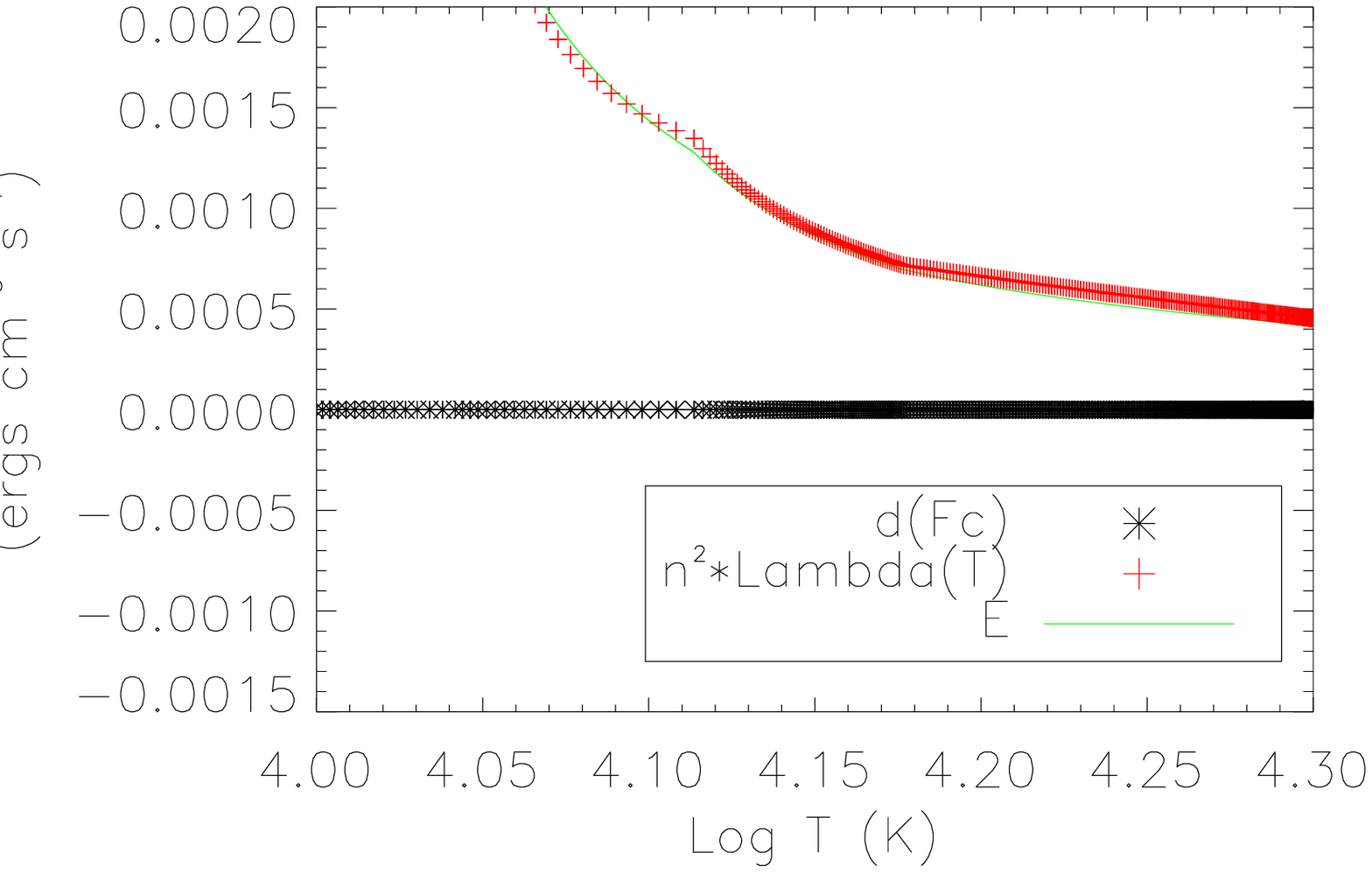}
\includegraphics[clip=true,trim=0  0 50 0,width=6.5cm]{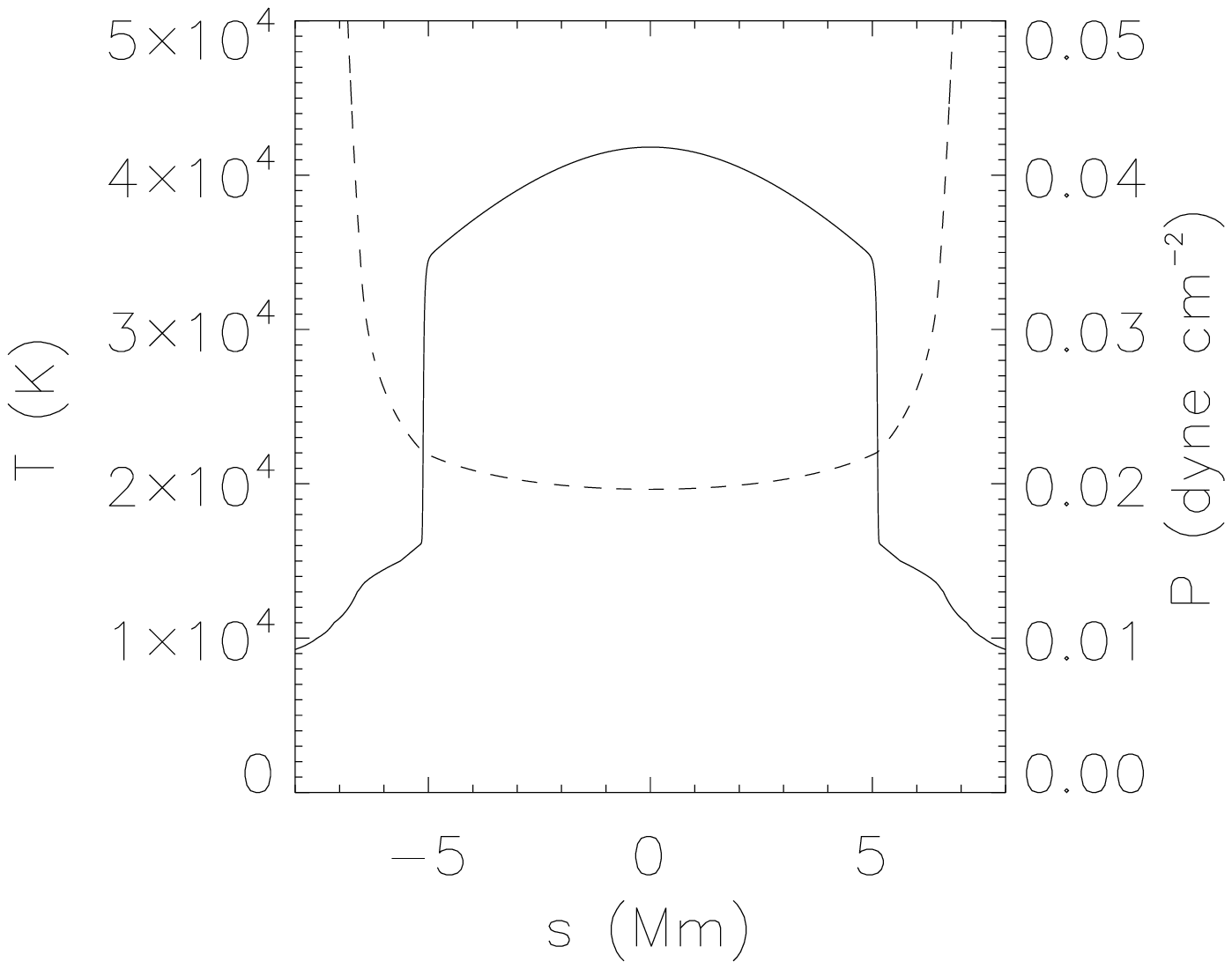}
\includegraphics[clip=true,trim=0  0 30 0,width=6.5cm]{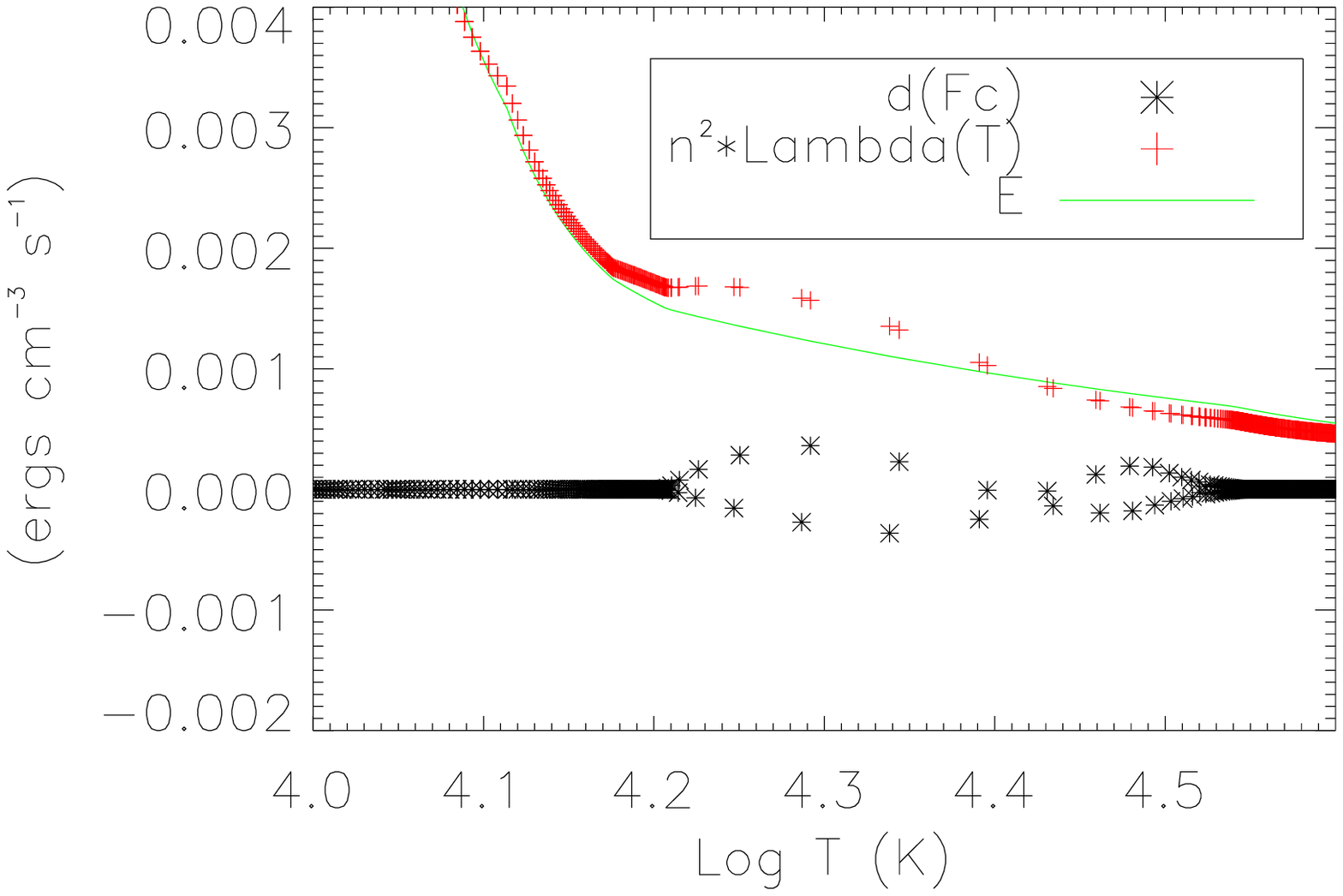}
\caption{Top panel, left: temperature (solid line) and pressure (dashed line) 
as a function of the curvilinear coordinate along the field lines, $s$. Right: 
divergence of the conductive flux (asterisks), radiative losses (crosses) and 
heating rate, $E$ (solid line), as a function of the temperature, for loop 13. 
Middle and bottom panels: as in the top panels for loop 16 and 22, 
respectively.}
\label{fig:TP}
\end{figure*}

We perform new simulations with the radiative loss function
$\Lambda_\mathrm{kp}$, and use constant heating rate per particle, setting
$\gamma=1$ in Eq.~\ref{eq:heating}. We list in Table~\ref{tab:1} and discuss
below a representative selection of cool loops in quasi-static equilibrium
that we obtained in these conditions. All loops are obtained starting from
four different loops of Paper I (Loops 17, 24, 26, and 27; the initial loop
parameters are at the top of each loop group in Table~\ref{tab:1}), by
changing the value of the constant $E_\mathrm{h}$ and the radiative loss
function. We are able to obtain quasi-static cool loops with maximum
temperature between $\sim1.5$ and $6.2\times10^4$~K, using $E_\mathrm{h}$ in
the range $0.2-140\times10^{-4}$~ergs~cm$^{-3}$~s$^{-1}$. We are not able to
obtain cool loops with maximum temperature in the range $4.3\lesssim\log
T\lesssim 4.5$~K, since at those temperature, $a$, defined in eq.~\ref{eq:aT},
becomes lower than 1 (see Sec.~\ref{sec:notes}) due to the change of the slope
of $\Lambda_\mathrm{kp}$. The cool loops found have the properties analytically
predicted by \citet{an86}: they are small ($L/2 = 7.5-8.8$~Mm and $h =
1.67-3.05$~Mm), nearly isobaric, and in approximate balance between the
heating rate and radiative losses. They have also the low-pressure values in
the range predicted by \citet{an86}, even if some loops have higher pressure
(up to $\sim 7$ times) compared with the cool loops obtained in Paper I.  
\begin{table}
\caption{Cool loop parameters at the end of the simulations when the loops have
reached a quasi-static condition ($\gamma=1$). All loops are obtained starting 
from a loop of Paper I (the initial loop parameters are at the top of each
loop group).}
\label{tab:1}
\centering
\small
\begin{tabular}{cccccccc}
  \hline\hline
  Loop&$E_\mathrm{h}$&$T_\mathrm{max}$&$P$&$L/2$&$h$\\
  &$10^{-4}$~ergs~cm$^{-3}$~s$^{-1}$&MK&dyne cm$^{-2}$&Mm&Mm\\
  \hline  
  \multicolumn{6}{c}{Loop$_i$: 17} \\
        & 0.2  &0.242  &0.008  &7     &1.12  \\ 
  \hline  
  1     & 0.2  &0.015  &0.0003 &7.7   &1.90  \\ 
  2     & 1    &0.042  &0.0008 &8.3   &2.50  \\ 
  3     & 4    &0.062  &0.002  &8.8   &3.05  \\ 
  \hline  
  \multicolumn{6}{c}{Loop$_i$: 24} \\
        & 6    &0.012   &0.024  &5.2   &0.27  \\ 
  \hline  
  4     & 6    &0.017   &0.011  &7.6   &1.77  \\ 
  5     & 7    &0.019   &0.012  &7.6   &1.84  \\ 
  6     & 8    &0.022   &0.013  &7.7   &1.89  \\ 
  7     & 15   &0.049   &0.012  &7.9   &2.09  \\ 
  8     & 30   &0.053   &0.043  &8.1   &2.33  \\ 
  9     & 35   &0.055   &0.049  &8.2   &2.41  \\ 
  10    & 50   &0.057   &0.067  &8.3   &2.52  \\ 
  11    & 60   &0.058   &0.079  &8.3   &2.52  \\ 
  12    & 70   &0.059   &0.090  &8.4   &2.56  \\ 
  13    & 100  &0.061   &0.13   &8.4   &2.56  \\ 
  14    & 130  &0.059   &0.17   &7.6   &1.84  \\ 
  15    & 140  &0.058   &0.18   &7.7   &1.89  \\ 
  \hline
  \multicolumn{6}{c}{Loop$_i$: 26} \\
        & 7.4  &0.050   &0.026  &5.5   &0.32  \\ 
  \hline  
  16    & 7.4  &0.020   &0.012  &7.7   &1.86  \\ 
  17    & 30   &0.054   &0.042  &8.2   &2.32  \\ 
  18    & 50   &0.057   &0.067  &8.3   &2.52  \\ 
  \hline
  \multicolumn{6}{c}{Loop$_i$: 27} \\
        & 6    &0.087   &0.024  &2.3   &0.04  \\ 
  \hline  
  19    & 6    &0.016   &0.012  &7.5   &1.67  \\ 
  20    & 9    &0.038   &0.015  &7.7   &1.89  \\ 
  21    & 12   &0.042   &0.020  &7.8   &1.99  \\ 
  22    & 20   &0.051   &0.030  &8.0   &2.20  \\ 
  23    & 25   &0.052   &0.037  &8.1   &2.28  \\ 
  24    & 28   &0.053   &0.040  &8.1   &2.32  \\ 
\end{tabular}
\end{table}
In Fig.~\ref{fig:TP} we plot the behavior of the loop parameters as well as 
of the terms of the energy equation for three loops 
chosen as examples (loops 13, 16 and 22 from top to bottom) at the end of the
simulation. The left panels show the temperature (solid line) and the pressure
(dashed line) profiles as a function of the curvilinear coordinate, $s$, while
the right panels show the radiative losses energy term, $n^2\Lambda$
(crosses), the heating rate, $E$ (solid line), and the divergence of the
conductive flux, $\nabla \mathbf{F}_{\textmd{\tiny{c}}}$ (asterisks), as a
function of the temperature. For all the loops, the pressure is constant along
the loop (within 1\% above the chromosphere) and the terms $n^2\Lambda$ and $E$
are in approximate balance, while the divergence of the conductive flux is
only a small term. From the left panels of Fig.~\ref{fig:TP} we see that the
temperature of the loops starts to increase slowly till a certain value of $s$
and then increases rapidly till the maximum temperature value reached by the
loop. The values of $L/2$ in Table~\ref{tab:1} for these loops include the 
piece where the temperature rises slowly.
\begin{table}
\caption{As in Tab.~\ref{tab:1} for intermediate-temperature loops ($\gamma=1$).}
\label{tab:2}
\centering
\small
\begin{tabular}{cccccccc}
  \hline\hline
  Loop&$E_\mathrm{h}$&$T_\mathrm{max}$&$P$&$L/2$&$h$\\
  &$10^{-4}$~ergs~cm$^{-3}$~s$^{-1}$&MK&dyne cm$^{-2}$&Mm&Mm\\
  \hline  
  \multicolumn{6}{c}{Loop$_i$: 17} \\
        & 0.2  &0.242  &0.008  &7     &1.12  \\ 
  \hline  
  25    & 5    &0.206  &0.008  &8.9   &3.14  \\ 
  26    & 10   &0.431  &0.036  &9.2   &3.44  \\ 
\end{tabular}
\end{table}
\begin{figure*}
\centering
\sidecaption
\includegraphics[clip=true,width=6cm]{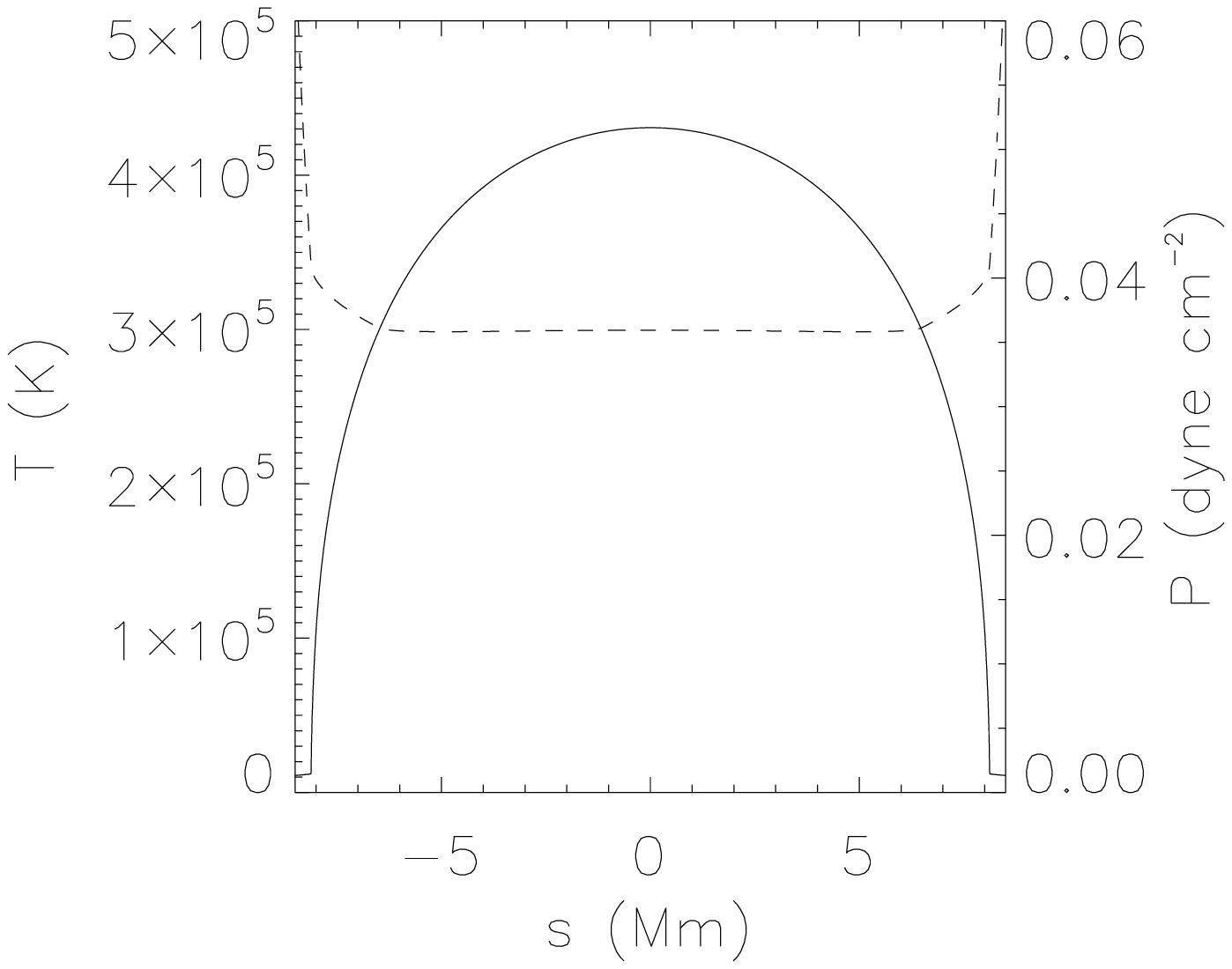}
\includegraphics[clip=true,width=6cm]{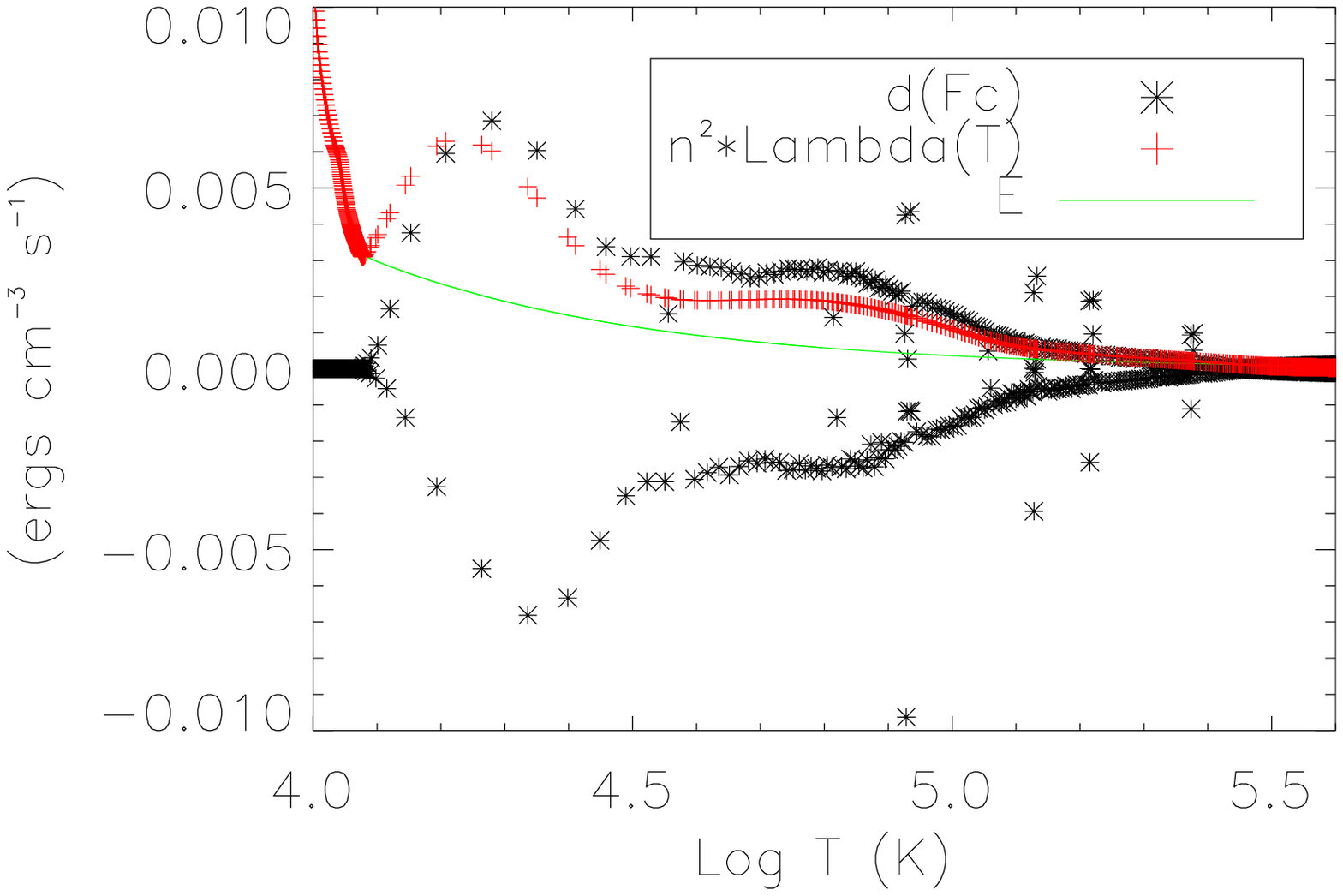}
\caption{Top panel, left: temperature (solid line) and pressure (dashed line) 
as a function of the curvilinear coordinate along the field lines, $s$. Right: 
divergence of the conductive flux (asterisks), radiative losses (crosses) and 
heating rate, $E$ (solid line), as a function of the temperature, for loop 26.}
\label{fig:TP2}
\end{figure*}

Using the same radiative loss function and $\gamma=1$, we also obtained
quasi-static intermediate-temperature loops ($0.1<T< 0.5 \times 10^5$~MK),
listed in Table~\ref{tab:2}. These loops are obtained by starting the
simulations from the quasi-static cool loop 17 of Table~1 in Paper I. For
loop 26, we show in Fig.~\ref{fig:TP2} the behavior of the temperature and the
pressure as a function of $s$ (left) and of the terms of the energy equation
as a function of the temperature (right). The divergence of the conductive
flux, comparable to the radiative losses, contributes to dissipate the heating
in excess.

\subsection{Relations between loop parameters and scaling laws}

\begin{figure*}
\centering
\sidecaption
\includegraphics[clip=true,width=12cm]{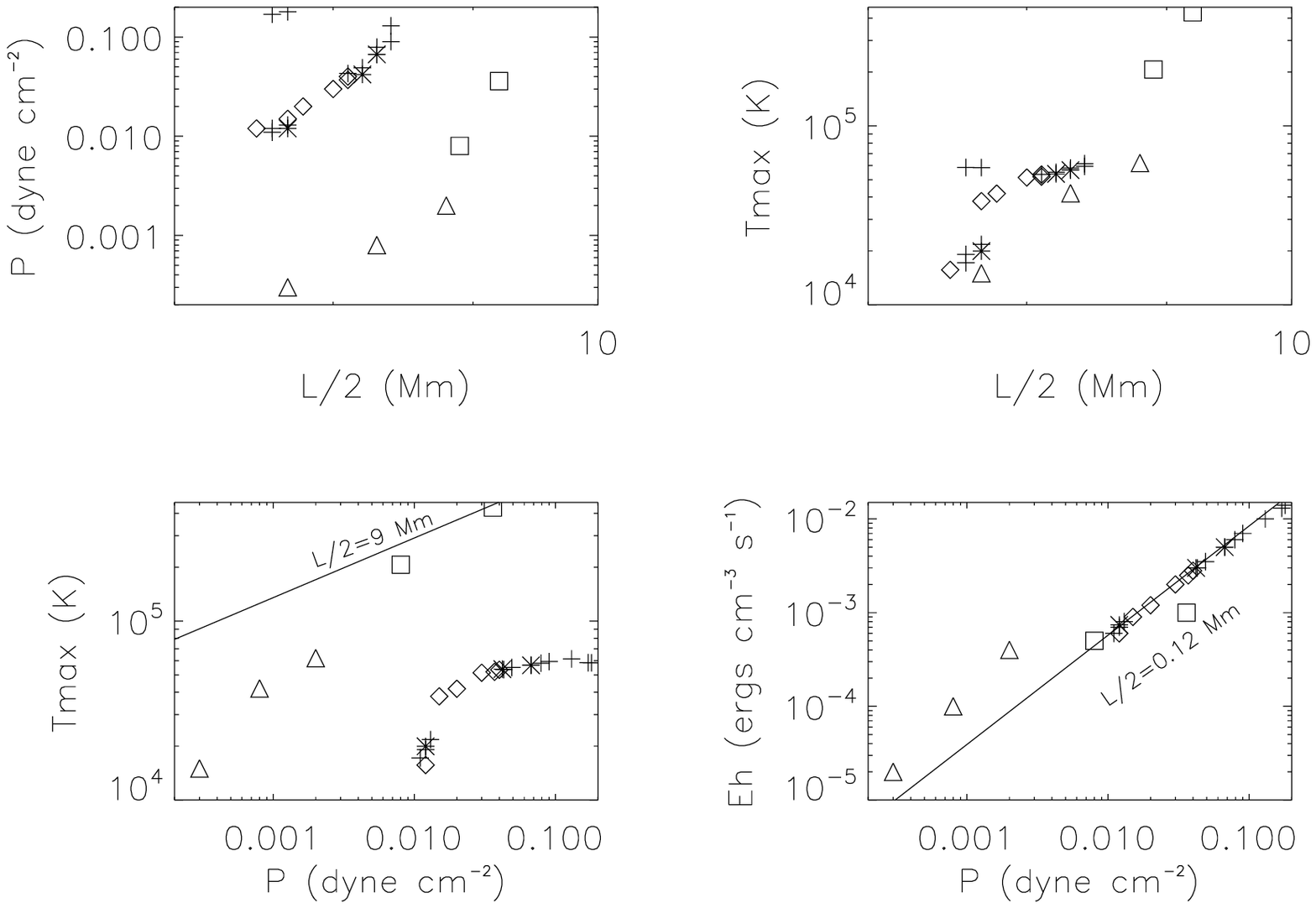}
\caption{Behavior of the physical parameters for loops 1–3 (triangles), 4–15
(crosses), 16–18 (asterisks), and 19-24 (diamonds) of Table~\ref{tab:1}, and
loops 25--26 (squares) of Table~\ref{tab:2}. The solid lines represent the RTV
scaling laws for coronal loops for different values of L/2.}
\label{fig:leggiscala}
\end{figure*}

In Fig.~\ref{fig:leggiscala} we show the relations between the thermodynamic
parameters ($P$, $T_\mathrm{max}$ and $L/2$) and $E_\mathrm{h}$ for the
loops in Table~\ref{tab:1} (loops 1--3 are represented by triangles, loops
4--15 by crosses, loops 16--18 by asterisks, and loops 19--24 by diamonds) and
Table~\ref{tab:2} (represented by squares). The solid lines in the lower
panels of Fig.~\ref{fig:leggiscala} represent the ``static'' scaling laws for
coronal loops described by \citet[hereafter RTV]{rtv} for different values of
$L/2$. The pressure of all cool loops with $T<0.1$~MK is proportional to
$E_\mathrm{h}$ and it is dependent on their length and maximum
temperature. In the simulations, indeed, we increase $E_\mathrm{h}$ in order
to have higher temperature loops, obtaining also higher pressure and
longer loops. There is, however, a maximum limit of $E_\mathrm{h}$
(different for each initial condition-loop) at which, even increasing its
value, the loops continue increasing their pressure but not their maximum
temperature (see bottom-left panel of Fig.~\ref{fig:leggiscala}).   

Intermediate-temperature loops obey the RTV scaling law for coronal loops for
temperature and pressure (see bottom-left panel of
Fig.~\ref{fig:leggiscala}) as the intermediate-temperature loops we found in
Paper I. Observed intermediate-temperature loops do not obey the coronal
scaling laws \citep{brown}, but as we already discussed in Paper I, the static
model used by \citet{rtv} to derive the relationships between coronal
temperature, pressure, length and heating in coronal loops does not seem to
accurately predict the physical conditions of these loops. We obtained
intermediate-temperature loops with pressures that are 1--2 orders of
magnitudes lower than measured in observed loops with the same temperatures
\citep{brown}. Loops 25--26 have the right pressures to fall on the scaling
laws lines.

\subsection{Calculated DEMs for cool and intermediate-temperature 
loops}\label{sec:dem}

\begin{figure}
\centering
\includegraphics[clip=true,width=9cm]{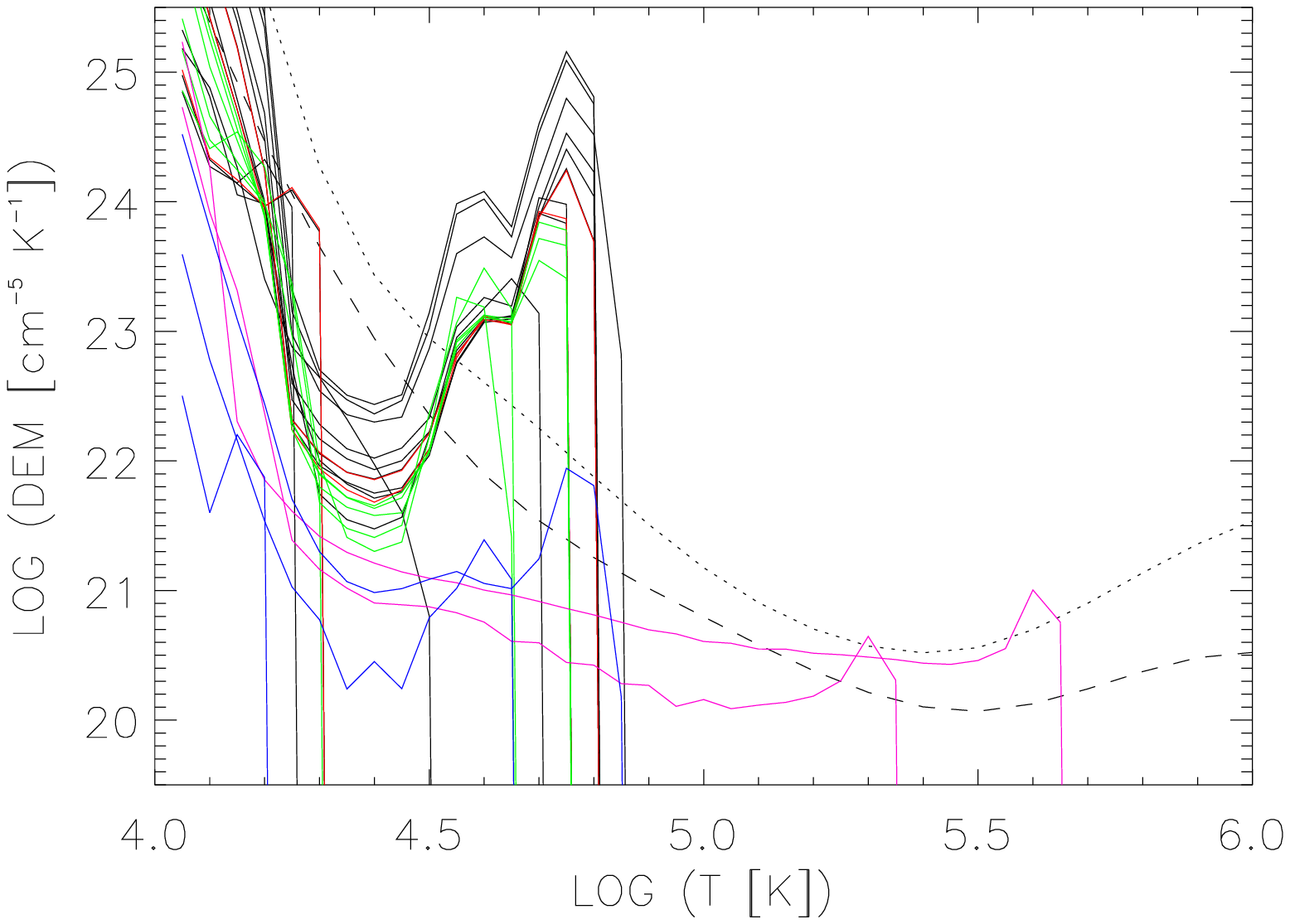}
\includegraphics[clip=true,width=9cm]{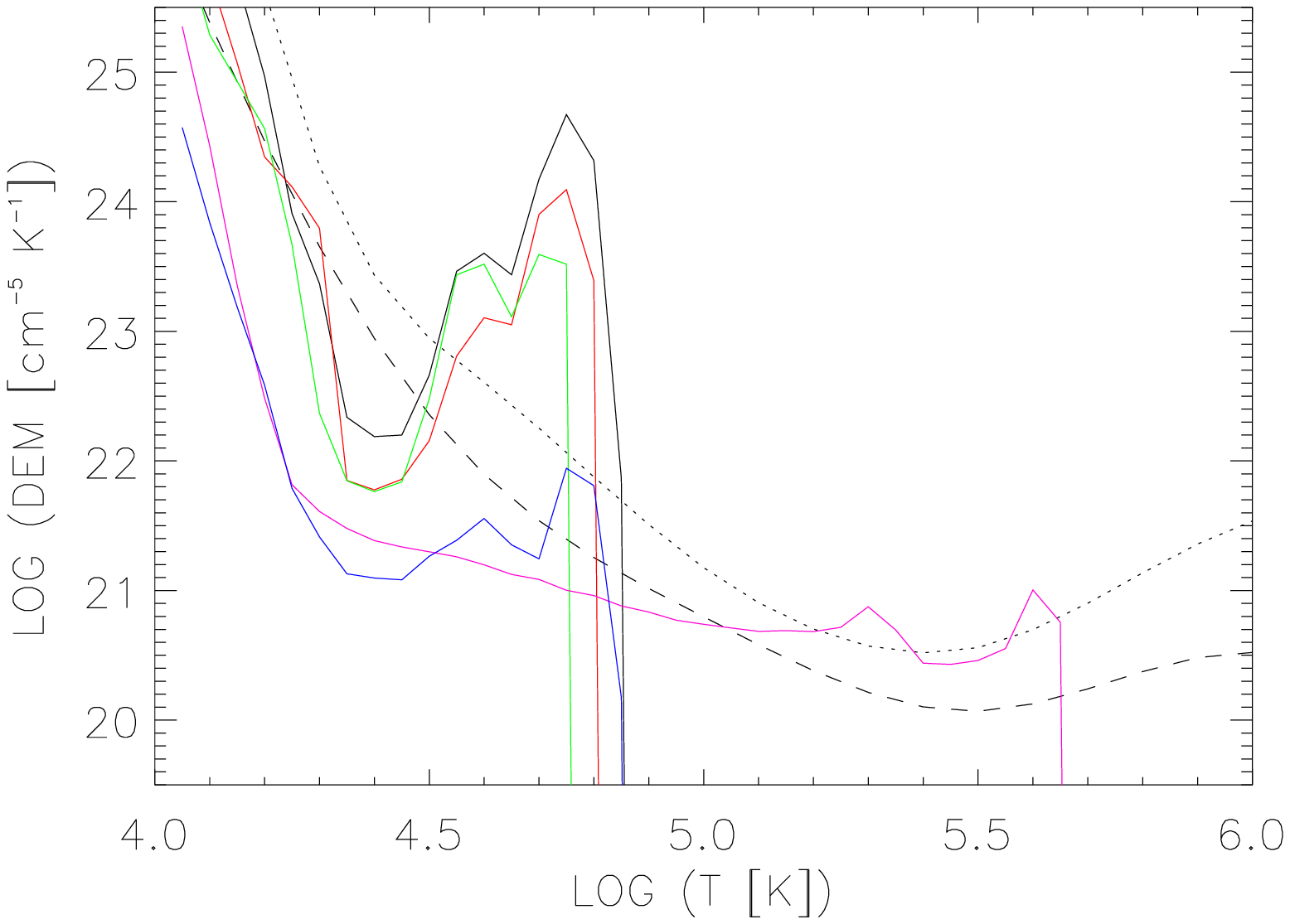}
\caption{Top: Calculated DEMs for the quasi-static cool loops 1--3 (solid 
blue lines), 4--15 (black), 16--18 (red), 19--24 (green) of Table~\ref{tab:1},
and the intermediate-temperature loops 25--26 (magenta) of Table~\ref{tab:2},
compared to the DEMs of the quiet Sun (dashed) and active region (dotted) from
the ``CHIANTI'' atomic data base \citep{chianti}. Bottom: total DEMs for each
group of loops shown in the top panel.}
\label{fig:dem1}
\end{figure}
\begin{figure}
\centering
\includegraphics[clip=true,width=9cm]{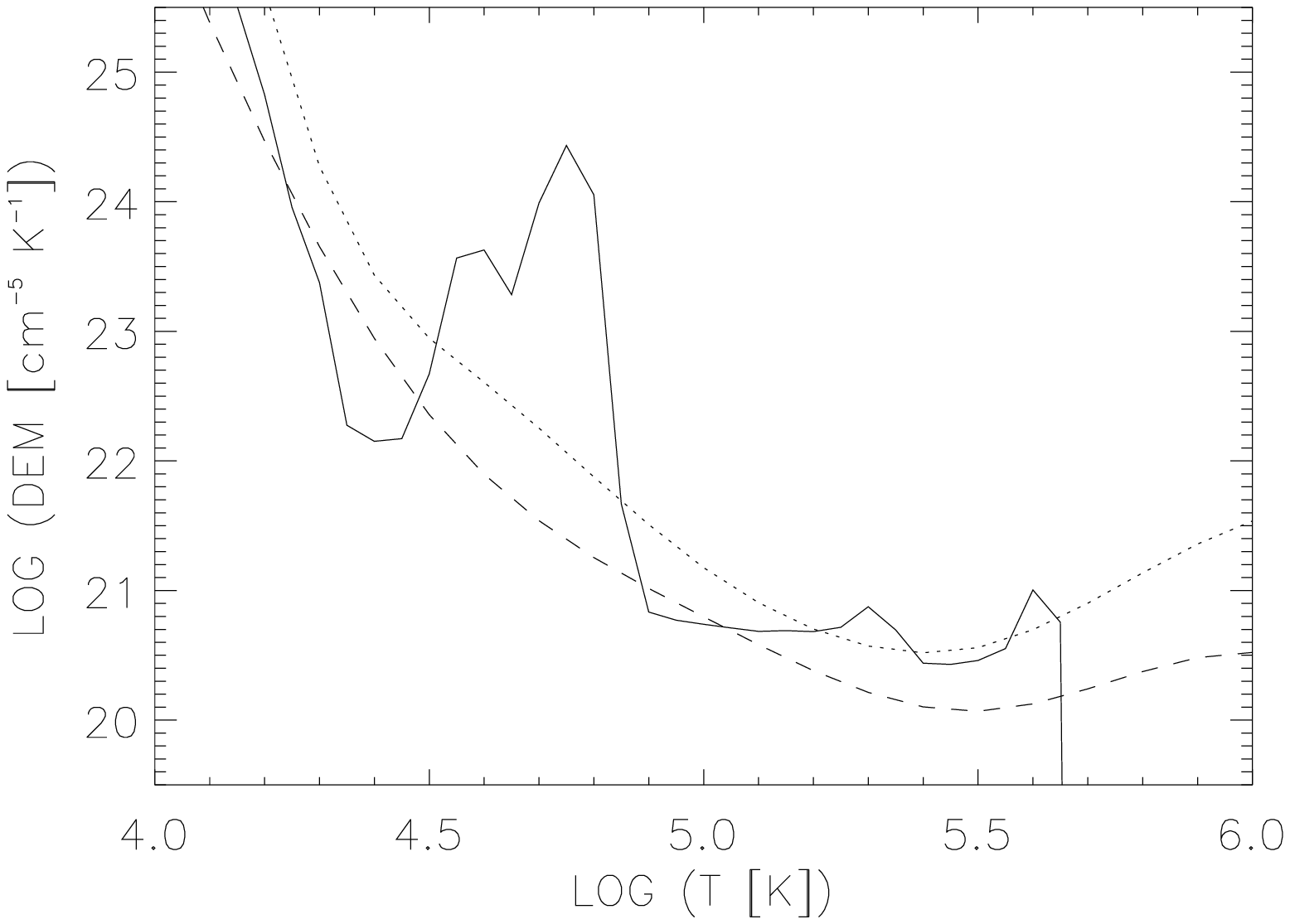}
\caption{Total DEM resulting from the combination of the DEMs of the loops 
1--24 and 25--26 (solid line), compared to the DEMs of the quiet Sun (dashed) 
and active region (dotted) from the ``CHIANTI'' atomic data base \citep{landi}.}
\label{fig:dem2}
\end{figure}

The theoretical DEMs for a single or isolated loop were computed for the
quasi-static loops we found according to \citet{spadaro}, with a temperature bin 
of 0.05~dex on a $\log T$ scale along the loop:
\begin{eqnarray}
DEM=n^2\frac{ds}{dT}.
\end{eqnarray}

This simplified approach permits to study the overall properties of
the DEM of this class of loops without the need for taking into account
details such as the shape of the loop, the geometry of the observations,
the loop cross-section, etc.

The upper panel of Fig.~\ref{fig:dem1} shows the calculated DEMs versus
temperature of the quasi-static cool loops 1--3 (solid 
blue lines), 4--15 (black), 16--18 (red), 19--24 (green) of Table~\ref{tab:1},
and the quasi-static intermediate-temperature loops 25--26 (magenta) of
Table~\ref{tab:2}. In this figure and in the next ones, we plot, for
comparison, the observed DEMs of the quiet Sun and active region (dashed and
dotted lines, respectively), derived using the \citet{vernazza} average quiet
Sun and active region intensities, and produced as part of the ``CHIANTI''
atomic data base collaboration \citep{landi}. In the lower panel of
Fig.~\ref{fig:dem1} we plot the total theoretical DEMs for each group of loops
obtained starting from a different initial loop-condition (distinguished by
the different colors). Assuming that the loops are equiprobable (uniformly
distributed in $\log T$) and with the same cross-section, we divided the
temperature range into bins of amplitude $0.2$~dex on a $\log T$ scale, and
considered for each bin a representative loop, i.e. a loop whose maximum
temperature belongs to that bin (our loops are almost isothermal). The total
DEMs are obtained by summing the DEMs of these representative loops. When more
loops have their maximum temperature falling in the same bin, we averaged
their DEMs. 

Using $\Lambda_\mathrm{kp}$ (and $\gamma=1$), we obtain cool loops with maximum 
temperatures covering the temperature range up to the position of its 
peak ($\log T\sim 4.8$~K) except for the interval $4.3\lesssim\log T\lesssim
4.5$. Adding the DEMs of the intermediate-temperature loops 25--26, the
resulting DEM (black solid line in Fig.~\ref{fig:dem2}) follows the shape of
the observed ones, except for the interval $4.6\lesssim\log T\lesssim 4.8$,
where we have an excess of emission due to the high density of the loops with
maximum temperature falling in that interval. Only Loop 2 and 3, with maximum
temperature belonging to this interval, have a pressure such that their DEMs
would resemble the observed one, but the solar total pressures at these
temperatures and heights, according to the model of \citet{avrett} are
estimated around 0.1~dyne~cm$^{-2}$ that is much higher than the pressures of
loops 2 and 3 and closer to that of all other loops of
Table~\ref{tab:1}. Moreover, the presence of the Ly-$\alpha$ peak at
log$T\sim4.2$~K and, in particular, the negative slope of the radiative loss
function (as explained in Sec.~\ref{sec:notes}), produces a relative minimum
in all DEMs, which remains in the total DEM (lower panel of
Fig.~\ref{fig:dem1}, blue, black, green or red lines). There are, however, in
the literature, derived quiet Sun DEMs \citep{dem1} that exhibit a minimum
around log$T\sim4.2$~K. 

There is a minimum in the total DEM also around log$T=4.9$~K that is caused by
the lack of cool loops with that maximum temperature. This minimum almost
corresponds to the maximum of the function $\Lambda_\mathrm{kp}$ or better to the
point where its slope starts to change and we have $a<1$. So, the lack of
cool loops with maximum temperature around $\log T=4.9$~K it is not caused by
an incomplete exploration of the parameter space but by the negative slope of
$\Lambda_\mathrm{kp}$ that prevents their formation. However, the shape of the
averaged DEM of the loops 25--26, with a flat minimum and a tail extended
towards low temperatures, helps filling this gap, improving the agreement with
the observed DEM. Since we considered a filling factor of $100\%$ the total
DEM has its highest value. With a lower filling factor the height of the DEM
would be lower.

\begin{figure}
\centering
\includegraphics[clip=true,width=9cm]{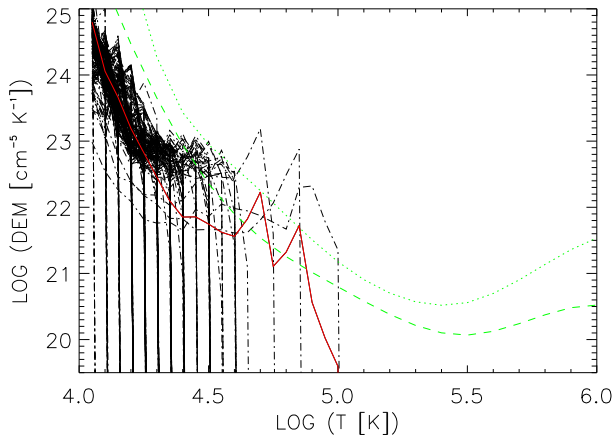}
\caption{Calculated DEM of Loop0 over the first 42 min of the simulation (red
  line). It is obtained by averaging the DEM of each loop obtained at each
  time step of the simulation (black dash-dotted lines). It is compared to the
  DEMs of the quiet Sun (green dashed line) and active region (green dotted
  line) from the ``CHIANTI'' atomic data base \citep{landi}.}
\label{fig:demloop0}
\end{figure}

We calculate also the emission due to Loop0, obtained in
Sect.~\ref{sec:heatingvolume} by performing a simulation using constant
heating rate per unit volume ($\gamma=0$) and starting from a quasi-static
cool loop of maximum temperature $T_\mathrm{max}\sim 1.2\times 10^4$~K. The
loop becomes a quasi-static ``coronal'' loop, after $\sim 2.5$~h from the
beginning of the simulation, reaching a top temperature of $\sim 8.5\times
10^5$~K. The evolution of Loop0's mean temperature, density and pressure is
shown in Fig.~\ref{fig:3}. 

The discussion that follows is based on considering each recordered step of
the simulation as a single dynamic loops at a particular instant of its
evolution (for example, cooling down or heating up depending if we keep the
heating on or we shut it down). We show in Fig.~\ref{fig:demloop0}, indeed,
the DEM of each loop obtained at each time step of the simulation (black
dot-dashed lines) during the first 42 min in which Loop0 evolves keeping its
temperature lower than $10^5$~K. The red line is the total DEM obtained by
combining all the loops as already explained. Another possible way to
calculate the total DEM is described in \citet{susino}. They simulated the DEM
of a multi-stranded loop by averaging instantaneous DEMs calculated at n
different times, randomly selected throughout the simulation. This approach is
based on the assumption that the states of the model at n randomly selected
times can be used to describe the behaviour of n independent strands observed
at the same time. In this analysis, we do not want to concentrate on how the
loops are obtained but we only want to show how the emission measure produced
by this particular distribution of loops looks like. The total DEM resembles
quite well the observed one and we do not have any of the problems observed
with the total DEM obtained from static loops. We are able to obtain also
loops with maximum temperature prohibitive for the quasi-static loops ($\log T
\sim 4.2$ and $\sim 4.9$~K). 

Obviously, the resulting DEM depends on the assumption we are making, in
particular on the number of the loops that fall in a certain temperature
interval and/or the filling factor and, obviously, it depends on the
distribution of the loops, and, ultimately, on the distribution of the heating
rates \citep{an86}.

\subsection{Non-equilibrium phase of Loop0}

We have furthermore examined the behaviour of the vertical component of
velocites in Loop0. At each time step of the simulation we computed the
mean value weighted by the DEM \citep[e.g.:][]{spadaro} of the vertical
component of velocities in temperature bins of 0.20~dex in $\log T$, considering
separately the two halves of the loop.  The 
temperature bins chosen are centered at $\log T=4.1$, $4.3$, $4.5$, $4.7$, and $4.9$. The last two bins are populated
only towards the end of the transient phase. We found that in the transient phase
we are considering, the vertical velocities are of the same magnitude and sign
at both footpoints.  

During the transient phase, the mean vertical component of velocities averaged on the whole loop for the different temperature bins appear in a few bursts lasting 1-4~minutes and reaching values of the order of
5-10~km/s or more in absolute value. After the first 10-15 minutes of the simulation,the velocities in these bursts are sistematically negative, adopting a sign convention which corresponds to negative Doppler shifts (redshifts). Considering the episodic character of these Doppler shifts, the average values
in each temperature bin over the 42~minutes interval correspond to redshifts of the order of
-1~km/s or less. These redshifts, however, are limited to the range of temperatures
covered by the transient phase (see Fig.~\ref{fig:demloop0}). At later times, 
as the loop reaches near coronal temperatures, the vertical velocities start
to oscillate
between blue- and red-shifts, with decreasing amplitudes until a quasi-static
situation is attained. \citet{peter1} report observed values of
about -5~km/s in the range $\log T =4.5 -5$, with one exception of nearly zero
wavelength shift. These redshifts are higher than our average redshifts, even though it should be noted
that there are only a few measurements in the temperature range best covered
by the transient phase of the simulations ($\log T < 4.6$). Their Table 3
lists only three lines nominally forming at or below $\log T=4.7$,
i.e. \ion{He}{i} 584 {\AA}, \ion{C}{ii} 1036 {\AA} and \ion{C}{ii} 1037 {\AA}. It is however
encouraging that our results for transient phase of Loop0 show a predominance
of redshifts, although this result should be confirmed and extended with
simulations spanning a variety of loop parameters.

\section{Conclusions}\label{sec:concl}

We have studied the conditions of existence and stability of cool loops with
$T\lesssim0.1$~MK through hydrodynamic simulations, introducing an
optically thick radiative loss function. We analyzed two different cases:
constant heating rate either per volume or per particle. We found that
it is possible to obtain quasi-static (velocities lower than $1$~km/s) cool
loops, as predicted by \citet{an86}, only by using a constant heating rate per 
particle, unlike the previous work in which we used different radiative loss
functions, with a less pronounced Ly-$\alpha$ peak.

We also obtained quasi-static loops with maximum temperature in the range
$1-5\times 10^5$~K, using the same optically thick radiative loss
function. These loops are smaller with respect to coronal loops but have
different characteristics compared to the static cool loops proposed by
\citet{an86} and others. They obey the scaling laws for coronal loops contrary
to results of previous works based on the observational data
\citep[e.g.,][]{brown}. The loops obtained have indeed low pressures that make
their parameters obey the RTV scaling laws, but these pressures are 1--2
orders of magnitudes lower than those estimated from observations
\citep{brown}. 

We examined and discussed the quasi-static solutions we found and analyzed
the contributions of the cool and intermediate-temperature loops to the TR
DEM, finding that a combination of these loops (assuming that they were
uniformly distributed), precisely because of their computed pressures, can
give a DEM with a shape not too far from the observed one for $\log T<4.3$ and
$\log T>5.0$. However there is a pronounced excess emission due to the high
density of the cool loops between $4.6\lesssim\log T_\mathrm{max}\lesssim 4.8$
and a deficit around $\log T \sim 4.4$ (see Sec.~\ref{sec:notes}).  

In this work we also showed a dynamic loop (Loop0), obtained by performing a
simulation using constant heating rate per unit volume and starting from
a quasi-static cool loop of maximum temperature $T_\mathrm{max}\sim 1.2\times
10^4$~K. The loop becomes a quasi-static ``coronal'' loop, after $\sim 2.5$~h
from the beginning of the simulation, reaching a top temperature of $\sim
8.5\times 10^5$~K. While the final state does not reproduce the observed
DEM for temperatures lower than $10^5$~K, the average DEM of Loop0,
interpreted as a combination of a set of evolving dynamic loops,
reproduces quite well the observed DEM. The whole simulation that we called
``Loop0'' can also be considered as the evolution of a single loop emerging
from lower atmospheric layers to the corona. The dimensions of this emerging
loop, its initial and final temperatures, and the time-scale of the event are
comparable to the observations and simulations of an emerging magnetic loops
from photosphere to low corona as the one described in the work of 
\citet{guglielmino}. 

In principle, cool and intermediate-temperature loops could be observed with
current telescopes, but in order to resolve them in all their temperature
extension, we would need multi-temperature observations, i.e. different UV
lines formed at temperatures between $0.01-1$~MK with resolution of at least
1''. Highly dynamical cool, low-lying loops have recently been reported by
\citet{hansteen} using observations obtained with the IRIS spacecraft
\citep{IRIS}. That kind of loops are usually observed as time-dependent,
short-lived ``segments'', not as complete loops. This could depend on the fact
that those loops extend over a range of temperatures not enterely covered by
the IRIS spectral lines. Such observations suggest that the class of loops
reported by \citet{hansteen} is related to short-lived, episodic heating;
``temporary'' loops would therefore be created and then rapidly
collapse. Hansteen et al. also stress that these are high-density structures
and postulate that these loops follow near-horizontal magnetic field (hence:
they are low-lying).

Based on the work of \citet{an86} and Paper I, we expect cool loops to be
low-lying even though we focuse our attention on steady-state heating. In this
work, we confirm that the existence, stability and properties of cool loops
strongly depend on the details of the radiative loss function. We also 
find that considering a more realistic function, the derived DEMs depart
from the observations (see Fig.~\ref{fig:dem1}). On the other hand, transient
loops, like Loop0, display characteristics which are appealingly closer to
observations. Note that this class of transient loops does not necessarily
imply impulsive heating. The similarity between the DEM of the transient phase
of Loop0 and the observed one, together with the new observations of dynamic
small scale structures on the Sun, suggest us to focus our attention to
simulate dynamic cool loops. This conclusion is reinforced by noting that this dynamical loop is
characterized in its non-equilibrium phase by the predominance of redshifts at its footpoints appearing in bursts of the order of -5 -- -10~km/s and in average 
of the order of -1~km/s or less over the 42 minutes of the transient. Redshifts of this magnitude could be marginally consistent with existing
spectroscopic observations of redshifts in the transition region in the relative low
temperature range best covered by the transient phase of the simulation ($\log T
< 4.6$).  The episodic nature of these red-shifts could be investigated by IRIS
time-resolved spectroscopic observations.  We also plan to further investigate
this intriguing result with more simulations spanning a wider range of loop
parameters.

In this perspective, an important point to consider is the effects of
partial ionization of hydrogen on the hydrodynamics of the loop plasma. The
equations for mass, momentum and energy conservation adopted in our work are
for a fully ionized hydrogen plasma. This assumption is well verified in
our cool loops, which are characterized by plasma pressures in the range
$10^{-2}-10^{-1}$~dyne~cm$^{-2}$, according to the calculations reported in
Table 3 of \citet{kuin}. Only in three cases the pressure in the loop is above
$10^{-1}$~dyne~cm$^{-2}$, resulting in a significant fraction of neutral
hydrogen just below 2$\times10^4$~K \citep[see][]{kuin}. Note that the
fraction increases and becomes important even at higher temperatures as the
pressure becomes higher. Since \citet{hansteen} stress that the episodically
heated loop they observe are high-density structures, the simulation of
dynamic cool loops should take into account the fraction of neutral hydrogen
in the hydrodynamic equations.

\begin{acknowledgements}

This work was supported by the ASI/INAF contracts I/013/12/0 for the program
"Solar Orbiter - Supporto scientifico per la realizzazione degli strumenti
METIS e SWA/DPU nelle fasi B2-C".

\end{acknowledgements}

\bibliographystyle{aa}

\end{document}